\documentclass[runningheads,a4paper]{llncs}

\usepackage[colorlinks,
            linkcolor=black,
            anchorcolor=black,
            citecolor=black
            ]{hyperref}

\usepackage{amssymb}
\usepackage{times}
\usepackage{epsfig}
\usepackage{graphicx}
\usepackage{amsmath}
\usepackage{amssymb}
\usepackage{multirow}
\usepackage[linesnumbered,ruled]{algorithm2e}
\usepackage[font=small,labelfont=bf]{caption}
\usepackage{subcaption}
\usepackage{epstopdf}
\usepackage[flushleft]{threeparttable}
\usepackage{float}
\usepackage{xcolor}

\begin{document}
\mainmatter  

\title{Geometric Brain Surface Network For Brain Cortical Parcellation}
\titlerunning{Deep Brain Parcellation Network}

%
%
\author{Wen Zhang, Yalin Wang}
\authorrunning{Zhang et al.}

\institute{School of Computing, Informatics and Decision Systems Engineering, \\Arizona State University,
Tempe, AZ, USA}

\toctitle{}
\tocauthor{Authors' Instructions}
\maketitle

\begin{abstract}
A large number of surface-based analyses on brain imaging data adopt some specific brain atlases to better assess structural and functional changes in one or more brain regions. In these analyses, it is necessary to obtain an anatomically correct surface parcellation scheme in an individual brain by referring to the given atlas. Traditional ways to accomplish this goal are through a designed surface-based registration or hand-crafted surface features, although both of them are time-consuming. A recent deep learning approach depends on a regular spherical parameterization of the mesh, which is computationally prohibitive in some cases and may also demand further post-processing to refine the network output. Therefore, an accurate and fully-automatic cortical surface parcellation scheme directly working on the original brain surfaces would be highly advantageous. In this study, we propose an end-to-end deep brain cortical parcellation network, called \textbf{DBPN}. Through intrinsic and extrinsic graph convolution kernels, DBPN dynamically deciphers neighborhood graph topology around each vertex and encodes the deciphered knowledge into node features. Eventually, a non-linear mapping between the node features and parcellation labels is constructed. Our model is a two-stage deep network which contains a coarse parcellation network with a U-shape structure and a refinement network to fine-tune the coarse results. We evaluate our model in a large public dataset and our work achieves superior performance than state-of-the-art baseline methods in both accuracy and efficiency.


\end{abstract}

\vspace{-0.7em}
\section{Introduction}


Brain cortical surface, delineating the shape of the cerebral cortex, is highly folded and consists of a large number of functionally and structurally different regions~\cite{dale1999cortical}. Identifying these regions demands a consistent parcellation map of the cortical surface among a group of subjects. Generally, a brain atlas would be defined beforehand according to an anatomical protocol and then parcellation schemes in the individual brain shall be drawn by following the given atlas. However, the conventional ways to do this such as manual labeling~\cite{klein2012101} or registration-based mapping~\cite{zhang2016functional} are inefficient due to the requirements of expert guidance and heavy computation. Therefore, an automatic pipeline which effectively generates a projection from the atlas to an individual brain surface is much needed. However, it is challenging because of the diversity of brain shapes owing to different genetic and environmental affection~\cite{paus2013environment}. As a result, it is extremely difficult to locate boundaries of the brain regions. It is a common belief that a plausible solution shall collectively analyze brain surface features, e.g. cortical thickness and area, together with their embedded geometry structures, e.g. local surface structure. 




Deep learning is a powerful tool for geometric feature learning on the non-Euclidean structured data, such as graphs and triangular meshes~\cite{bronstein2017geometric}. Efforts to generalize the conventional image-based convolutional operators to these irregular data can be categorized into spectral~\cite{kipf2016semi} and spatial~\cite{monti2017geometric} approaches. Here, we favor the latter approach to encode both intrinsic and extrinsic surface structures. Until recently, only a few studies applied deep models to process brain cortical surfaces~\cite{seong2018geometric,wu2018registration}. However, there are some limitations in previous work. For example, inputs of their models are feature patches based on the intrinsic structure of the brain surface while the extrinsic structures are ignored. Another drawback of the patch-based study is that it is unable to change patch size dynamically according to the complexity of local shapes. Besides, these methods require a time-consuming parameterization of brain surfaces, i.e. a regular metric in a sphere. However, deformation from the brain surface to a sphere might introduce distortions and sometimes is difficult to achieve if there are topological defects on the original surface, e.g. holes or triangle flips. In addition, existing deep models designed for brain parcellation require post-processing on the network outputs.


To overcome the aforementioned limitations, we propose an automatic and end-to-end deep brain parcellation network (DBPN). To fully describe brain surfaces, we use two surface-based convolution kernels, i.e. intrinsic and extrinsic kernels, to aggregate local vertex features. The intrinsic kernel uses a polar coordinate system depending on the geodesic distance while the extrinsic kernel uses a Cartesian coordinate system which is embedded in the local Euclidean space around each vertex. The convolutional operation in these two kernels is carried out by referring a continuous B-spline kernel~\cite{fey2018splinecnn} which is efficient to compute. We design a two-step learning framework for the brain parcellation. The first learning is a coarse parcellation built with a U-shape deep network and relies on both intrinsic and extrinsic geometry features. The second is a refinement learning built upon the coarse parcellation scheme. This network depends on the intrinsic kernel to remove disconnected clusters and smooth the regional boundaries. We test our model in a large dataset containing 101 normal subjects and manually-created labels. The experimental results demonstrate the effectiveness and efficiency of our model. The main contributions of this work are 4 folds: 1) proposing an end-to-end deep model for surface-based brain parcellation; 2) directly working on the original brain surfaces without a sphere parameterization; 3) using graph convolution with the intrinsic and extrinsic kernels to capture local geometry structures; 4) designing a two-step learning scheme to construct a smooth parcellation map.

\vspace{-0.7em}
\section{Methods}
In this section, we first introduce two surface convolution kernels which capture intrinsic and extrinsic geometrical properties of human cortical surfaces. After that, based on these kernels, a two-stage deep network is proposed. In the first stage, we design a U-shape network to gather clustering information in the deep layers. The vertex clustering pattern, to a large extent, relies on the surface extrinsic properties. Then, we construct the second stage with a refinement network. It removes cluster noises caused by the spatial down/up-sampling in the first stage. This network only uses the intrinsic surface kernel to guarantee the smoothness of parcellation scheme on the surface.

\vspace{-0.7em}
\subsection{Surface-Based Convolution Kernels}
A surface can be realized from two aspects, i.e., extrinsic and intrinsic structures, and they are geometrically associated. The extrinsic structure of a surface depends on a coordinate system in an embedding space of that surface. Therefore, different coordinate systems might lead to distinct expressions of the extrinsic structures. The most common coordinate system for meshes and point clouds is the Cartesian coordinate in the Euclidean space. In contrast, the intrinsic structure is independent of the embedding space, meaning it can be measured within a surface itself without any reference to a specific coordinate system. An intrinsic structure example is the geodesic distance between two points in a surface. To have a comprehensive awareness of a brain surface, both of these geometry structures are indispensable.


\begin{figure}[t]
\center
        \includegraphics[width=1\linewidth]{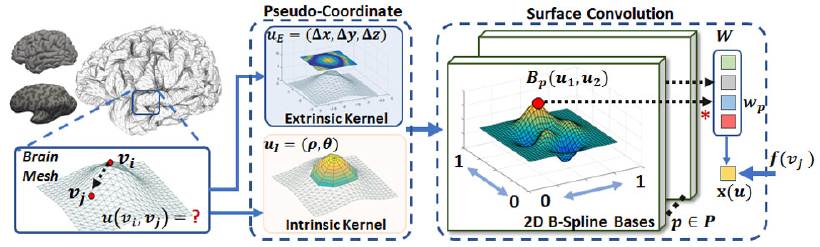}
    \caption{B-spline-based intrinsic and extrinsic convolution kernels. In these kernels, aggregation of local vertex features is regulated by the learned edge weight $x(u)$ which is determined by the intrinsic and extrinsic geometry structures, i.e., expressions of pseudo-coordinate $u(v_i,v_j)$. Here, we show an example of 2-dimensional B-Spline convolution (the right block). }
    \label{fig:bspline_kernel}
\vspace{-1em}
\end{figure}

A brain surface is generally represented as a local connected graph $G=\{V,E,X,F\}$, where $V=\{v_i\}^N_{i=1}$ is the set of nodes, i.e. vertices, $E=\{\epsilon_{i,j}\}$ is the set of edges and $X=\{x_{i,j}\}$ is the corresponding edge weights. Generally, $G$ is sparse meaning the number of edges $|E|\ll N^2$. A function $F=\{f(v_i)\in\mathbb{R}^M\}$ is defined on each vertex $v_i$, which can be regarded as node signals or features. The neighbourhood graph of $v$ is denoted by $\mathcal{N}(v)$. 

In this study, the convolution kernel encodes the local mesh geometry around a given node and meanwhile aggregates the neighboring node features such as cortical thickness and surface area. The output of this kernel is the new node features. We understand the aggregation mechanism as an analogy of image-based convolutional operation but differs on the definition of aggregation weights, i.e. $x_{i,j}$,  which are defined on edges rather than nodes. The aggregation is mathematically formulated as:
\begin{equation}\label{eq:agg}
    AGG(v_i)=\delta(\frac{1}{|\mathcal{N}(v_i)|}\sum_{v_j\in\mathcal{N}(v_i)}x_{i,j}*f(v_j)*W),
\end{equation}
where $W\in \mathbb{R}^{M_{in}\times M_{out}}$ is a trainable weight matrix controlling the feature dimension and $\delta$ is a non-linear activation. In Eq.~\ref{eq:agg}, we want $x_{i,j}$ to be a function based on the local geometry properties on edges via the pseudo-coordinates~\cite{bronstein2017geometric}, $u(v_i,v_j)\in \mathbb{R}^k$, and thus kernel value $x(u)$ is learned dynamically (Fig.~\ref{fig:bspline_kernel}). For the extrinsic geometric property, we use a 3-dimensional Cartesian coordinate system to reflect a shift from a source node $v_i=(x_i,y_i,z_i)$ to a target node $v_j=(x_j,y_j,z_j)$ in the Euclidean space, as $u_E(v_i,v_j)=(x_i-x_j,y_i-y_j,z_i-z_j)$. For the intrinsic geometric property, we use a 2-dimensional polar coordinates denoted by $u_I(v_i,v_j)=(\rho, \theta)$, where $\rho$ is the geodesic distance between $v_i$ and $v_j$, $\theta$ is the rotation angle in the tangent plane with respect to the maximal curvature direction. Noting that the pseudo-coordinates will scaled to a range [0,1] in each dimension. 

Since the brain surface is geometrically complex and its shape is diverse subject-to-subject, optimal $x(u)$ fitting to various geometry patterns is hard to specify. In this study, we use a spline-based convolution kernel~\cite{fey2018splinecnn} to automatically encode it. The idea is to generate a set of feature manifolds as B-spline surfaces and learn their combination weights as parameters. Mathematically, suppose that the pseudo-coordinates used in this kernel is $k$-dimensional, i.e. $k=$2 or 3, we define $k$ B-spline bases of degree $m$ as $(\{N_{i,j}^m\}_{j=1}^{d_i})_{i=1}^k$, based on equidistant knot vectors. $d_i$ is the kernel size defined in the $i$-th dimension of $u$. The larger $d_i$, the more control points in the B-spline surface. By referring the B-spline composition, we give the convolution kernel as:
\begin{equation}\label{eq:splineConv}
    x(u)=\sum_{p\in \mathcal{P}}w_p *B_p(u).
\end{equation}
$w_p$ is a trainable control weight for each element $p$ from the Cartesian product $\mathcal{P}={N_{1,i}^m}\times...\times{N_{k,i}^m}$ and $B_p(u)$ is the product of the basis functions, $B_p(u)=\prod_{i=1}^k N_{i,p_i}^m(u_i)$. We can interpret $B_p(u)$ as a B-spline surface in the feature space depend on $u(v_i,v_j)$ and the corresponding $w_p$ controls the weight of this surface. There are a total of $D=\prod_{i=1}^k d_i$ such kind of surfaces in $\mathcal{P}$ for each $u$ and thus the computation has a fixed time complexity for searching the $B_p(u)$ value in a B-spline surface.   

\begin{figure}[t]
\center
        \includegraphics[width=0.95\linewidth]{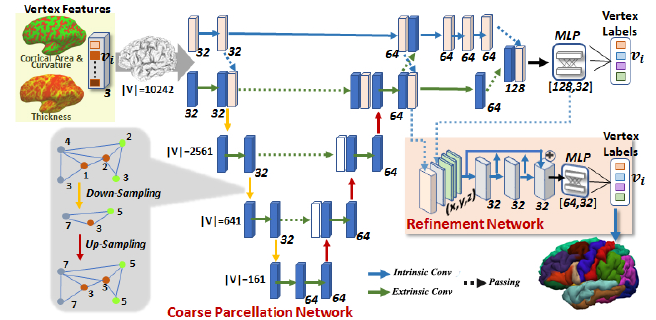}
    \caption{Pipeline of DBPN. We design the extrinsic and intrinsic convolutional kernels based on B-spline functions to encode local geometry patterns and aggregate node features. With these kernels, a two-stage parcellation is proposed. In the coarse parcellation network, intrinsic and extrinsic encoding is conducted concurrently, which accommodates a U-shape structure with graph spatial down/up-samplings. The refinement network further modifies the coarse results with intrinsic kernels. DBPN learns a projection from the vertex-wise features to the parcellation labels.}
    \label{fig:pipeline}
\vspace{-1em}
\end{figure}

\vspace{-0.7em}
\subsection{Coarse Parcellation Network}
This network serves as a coarse parcellation of brain surfaces, which projects 3 vertical features, e.g. surface area, mean curvature, and cortical thickness, to the label space. We have two parallel sub-networks (Fig.~\ref{fig:pipeline}). One uses the intrinsic convolution kernel to aggregates local vertex features up to k-hops in the surface graph. The other is built with the extrinsic convolution kernel and has a U-shape structure for node clustering. 

The U-net structure has a good performance in brain segmentation tasks~\cite{ronneberger2015u}. For the U-shape structure in our model, two key components are down-sampling and up-sampling operations on the surface graph. The down-sampling operation pools similar vertices together by searching meaningful neighborhoods on graphs. The pooled patterns are stored for the later up-sampling process which restores the original vertex order. Since clustering on the general graph is NP-hard, here, we use a multi-level graph clustering algorithm, called Graclus~\cite{dhillon2007weighted}, to approximate the graph clustering. This method has been proved to be effectively dealing with large graphs with thousands of vertices. It avoids eigenvector computation which is painful and prohibitive for large graphs but, instead, directly optimizes a weighted graph clustering objective, e.g. normalized cut. Each round of clustering groups vertices by pairs. The node value in the coarsened graph is the sum of grouped nodes while the new pseudo-coordinate is their average. The edge is the union of connection patterns (shown in the gray block, Fig.\ref{fig:pipeline}). For each down-sampling layer in our model, we perform graph clustering twice that divides the number of vertices by four. The up-sampling works in the opposite way by reversing the coarsening scheme. During this reverse operation, the value of a vertex in the coarser level will be passed back to and shared by all of its elements in the finer level. It is worth noting that we only deploy the U-shape network in the extrinsic sub-network. Since each of the down-sampling and up-sampling operations would change local graph topology, it is time-consuming to recompute the intrinsic pseudo-coordinates but the extrinsic kernels. 

In the end, we combine node features generated from the intrinsic and extrinsic sub-networks respectively and feed them into a multilayer perceptron (MLP) with a softmax in the last layer. The output of this coarse parcellation network is the probability of each vertex belonging to the given parcellation label. The object function of this network is the negative log-likelihood which is extensively utilized in multi-label classification.

\vspace{-0.7em}
\subsection{Refinement Parcellation Network}
The output of coarse parcellation might contain several discontinued clusters due to the down-sampling on the local extrinsic geometry rather than intrinsic structures. Therefore, we create a refinement network based on the surface intrinsic structure to removes noises in each parcellated region and meanwhile fine-tune the boundary of those regions. We add two additional vertex-wise information to the feature vector generated before the last softmax layer in the coarse parcellation network. One additional information is a copy of encoded vertex features in the middle of the intrinsic sub-network. The other is the vertex spatial coordinate vector $(x,y,z)$. In this refinement network, we add a skip method~\cite{xu2018representation} which let the network dynamically adopt various sizes of the neighborhood in aggregation. After 3 intrinsic convolutional layers, we use an MLP to obtain the vertex-wise probability. Here, in addition to the negative log-likelihood, we add n multi-class Dice score weighted by $\lambda$ to the vertex-wise loss function.
\vspace{-0.7em}
\begin{equation}\label{Eq:PredLoss}
\vspace{-0.7em}
\begin{split}
    D(g,\Tilde{g})=\frac{1}{|\mathbf{L}|}\sum_{l\in \mathbf{L}}\frac{2\sum_ig_l^i\Tilde{g}_l^i}{\sum_i(g_l^i+\Tilde{g}_l^i)}, 
\end{split}
\end{equation}
where $\{g_l^i\}_{i\in Y,l\in \mathbf{L}}$, $\{\Tilde{g}_l^i\}_{i\in Y,l\in \mathbf{L}}$ are the set of label probability vectors for all vertices for the ground truth and the prediction label. $|\mathbf{L}|$ is the number of brain regions.

\begin{figure}[t]
\center
        \includegraphics[width=0.9\linewidth]{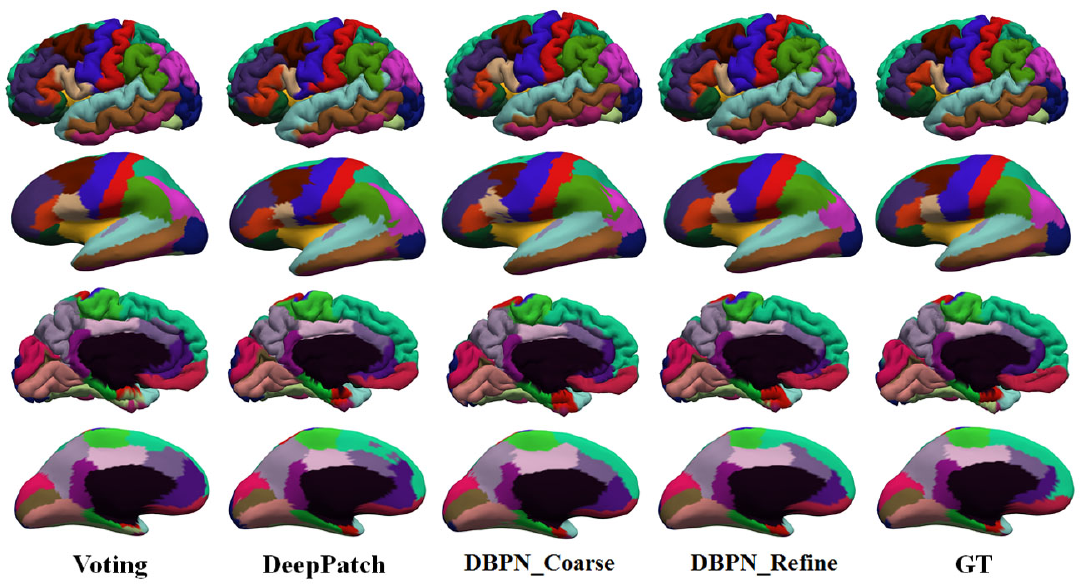}
    \caption{Visualization of brain surface parcellation. Along with the baseline methods, i.e., Voting and DeepPatch, we present our coarse (DBPN\_Coarse) and refined (DBPN\_Refine) parcellation results and compare them with the ground truth (GT) data.}
    \label{fig:visualresult}
\vspace{-1em}
\end{figure}

\vspace{-1em}
\section{Experimental Results}
\vspace{-0.7em}
\subsection{Dataset and Settings}
We evaluate our model on a large public dataset\footnote{https://mindboggle.info/data.html}, named Mindboggle-101, which contains a surface-based brain atlas drawn by following the Desikan–Killiany–Tourville protocol~\cite{klein2012101}. Only the left brain surfaces are evaluated and each surface is manually parcellated into 32 sub-regions. The brain surface is reconstructed by using the standard computation pipeline in FreeSurfer package\footnote{http://surfer.nmr.mgh.harvard.edu/} and 3 vertex-wise structural features, i.e. cortical thickness, surface area, and curvature, are measured. The 5-fold cross-validation is carried out in this study. Since each subject brain contains 10424 vertices, the total training samples are 833920 vertices (from 80 subjects). To assess model performance, we compute the average regional Dice score for each subject and report statistics for the group of testing. The hyper-parameter $\lambda$ in the refinement network is empirically set to 10. For the two-stage learning, we first train the coarse parcellation network with a learning rate $r=0.01$ and then reduce it to $r=0.0001$ when training the refinement network which has a learning rate $r=0.005$. The strategy of learning decay is added. 

Here, we pick two baseline methods for brain cortical parcellation. The first one, termed as Voting, is a multi-atlas based parcellation method which use majority voting to combine parcellation scheme in the training data~\cite{klein2012101}. Specifically, surface-based registrations project label information in the training data to the surface of the testing data. Then we do majority voting vertex-wisely to determine parcellation scheme in each subject and compute the statistics. The second one, termed as DeepPatch, is a deep learning model based on multi-channel feature images in local intrinsic patches~\cite{wu2018registration}. We extract vertex patches based on the local polar coordinate in the spherical domain and follow the network settings reported in the paper. Note that we do not include the post-process to remove noise in this method to have a fair comparison as a deep model.  

\begin{figure}[t]
\center
        \includegraphics[width=1\linewidth]{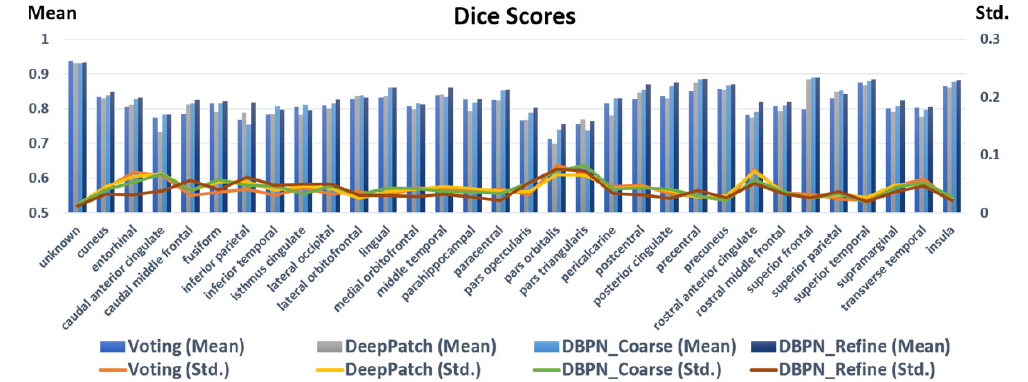}
    \caption{Comparison of brain regional Dice scores.}
    \label{fig:dice}
\vspace{-1em}
\end{figure}

\vspace{-0.7em}
\subsection{Results}
We show the parcellation scheme of a randomly selected sample in Fig.~\ref{fig:visualresult}. Both of the original and inflated brain surfaces overlaid with predicted labels are presented. Our method outperforms other baselines with the closest consistency to the ground truth and achieves a Dice score of $0.846\pm 0.034$ after refinement and $0.831\pm 0.037$ in the coarse parcellation, while the method with majority voting obtains $0.817\pm 0.038$ and the deep model with spherical patches achieves $0.821\pm 0.037$. Besides, we observe that, compared with the result from the coarse parcellation network, refinement network prominently remove discontinued clusters and improve the boundary prediction by adding smoothness. When considering the computational efficiency, our DBPN significantly outperforms the others. Since our method is an end-to-end framework, once the model has been well trained, testing on a subject will complete in a few seconds. In contrast, due to the burden computation of surface-based registration or spherical parameterization, other baseline methods might need hours to process.  

To further quantitatively address parcellation results in details, we compute the Dice score for each of the 32 subregions and report the statistics in Fig.~\ref{fig:dice}. Compared to majority voting method and the deep model with spherical patches, DBPN achieves better results in 27 out of 32 brain subregions and most of them have a clear boundary locating at the stable cortical landmarks, i.e. gyrus and sulci.

\vspace{-0.7em}
\section{Conclusion and Future Works}
We describes a novel surface-based deep learning model for brain parcellation. Two graph convolution kernels is designed based on the intrinsic and extrinsic geometric properties. Built on these kernels, a two-stage network parcellates cortical surface to a coarse map and then refines it. There are several interesting directions that are warranted for further investigation. For example, vertex features extracted from multimodal neuroimaging data can potentially improve parcellation accuracy. 

\vspace{-0.7em}
\subsubsection*{Acknowledgments.}
The research was supported by NIH (R21AG049216, RF1AG0517-10, R01EB025032, and U54EB020403). We gratefully acknowledge the support of NVIDIA Corporation with the donation of the Titan Xp GPU used for this research.

\bibliographystyle{splncs03}

\vspace{-1em}
\bibliography{fbs}

\begin{thebibliography}{10}
\providecommand{\url}[1]{\texttt{#1}}
\providecommand{\urlprefix}{URL }

\bibitem{bronstein2017geometric}
Bronstein, M.M., Bruna, J., LeCun, Y., Szlam, A., Vandergheynst, P.: Geometric
  deep learning: going beyond euclidean data. IEEE Signal Processing Magazine
  34(4),  18--42 (2017)

\bibitem{dale1999cortical}
Dale, A.M., Fischl, B., Sereno, M.I.: Cortical surface-based analysis: I.
  segmentation and surface reconstruction. Neuroimage  9(2),  179--194 (1999)

\bibitem{dhillon2007weighted}
Dhillon, I.S., Guan, Y., Kulis, B.: Weighted graph cuts without eigenvectors a
  multilevel approach. IEEE Trans Pattern Anal Mach Intell  29(11),  1944--1957
  (2007)

\bibitem{fey2018splinecnn}
Fey, M., Eric~Lenssen, J., Weichert, F., M{\"u}ller, H.: Splinecnn: Fast
  geometric deep learning with continuous b-spline kernels. In: CVPR. pp.
  869--877 (2018)

\bibitem{kipf2016semi}
Kipf, T.N., Welling, M.: Semi-supervised classification with graph
  convolutional networks. arXiv preprint arXiv:1609.02907  (2016)

\bibitem{klein2012101}
Klein, A., Tourville, J.: 101 labeled brain images and a consistent human
  cortical labeling protocol. Frontiers in neuroscience  6,  171 (2012)

\bibitem{monti2017geometric}
Monti, F., Boscaini, D., Masci, J., Rodola, E., Svoboda, J., Bronstein, M.M.:
  Geometric deep learning on graphs and manifolds using mixture model cnns. In:
  CVPR. pp. 5115--5124 (2017)

\bibitem{paus2013environment}
Paus, T.: How environment and genes shape the adolescent brain. Hormones and
  behavior  64(2),  195--202 (2013)

\bibitem{ronneberger2015u}
Ronneberger, O., Fischer, P., Brox, T.: U-net: Convolutional networks for
  biomedical image segmentation. In: International Conference on Medical image
  computing and computer-assisted intervention. pp. 234--241. Springer (2015)

\bibitem{seong2018geometric}
Seong, S.B., Pae, C., Park, H.J.: Geometric convolutional neural network for
  analyzing surface-based neuroimaging data. Frontiers in Neuroinformatics  12,
  ~42 (2018)

\bibitem{wu2018registration}
Wu, Z., Li, G., Wang, L., Shi, F., Lin, W., Gilmore, J.H., Shen, D.:
  Registration-free infant cortical surface parcellation using deep
  convolutional neural networks. In: MICCAI. pp. 672--680. Springer (2018)

\bibitem{xu2018representation}
Xu, K., Li, C., Tian, Y., Sonobe, T., Kawarabayashi, K.i., Jegelka, S.:
  Representation learning on graphs with jumping knowledge networks. arXiv
  preprint arXiv:1806.03536  (2018)

\bibitem{zhang2016functional}
Zhang, W., Wang, J., Fan, L., Zhang, Y., Fox, P.T., Eickhoff, S.B., Yu, C.,
  Jiang, T.: Functional organization of the fusiform gyrus revealed with
  connectivity profiles. Human brain mapping  37(8),  3003--3016 (2016)

\end{thebibliography}

\end{document}